\documentclass[prl,aps,twocolumn,showpacs,preprintnumbers,amsmath,amssymb]{revtex4}
\def\lbldef#1#2{\expandafter\gdef\csname #1\endcsname {#2}}

\def\href#1#2{#2}

\newcommand{\beq}{\begin{equation}}
\newcommand{\eeq}{\end{equation}}
\newcommand{\ba}{\begin{eqnarray}}
\newcommand{\ea}{\end{eqnarray}}

\def\Box{\kern1pt\vbox{\hrule height 1.2pt\hbox{\vrule width 1.2pt\hskip 3pt
   \vbox{\vskip 6pt}\hskip 3pt\vrule width 0.6pt}\hrule height 0.6pt}\kern1pt}

\def\nn{\nonumber} \def\bd{\begin{document}} \def\ed{\end{document}}
\def\ds{\documentstyle} \let\fr=\frac \let\bl=\bigl \let\br=\bigr
\let\Br=\Bigr \let\Bl=\Bigl
\let\bm=\bibitem
\let\na=\nabla
\let\pa=\partial \let\ov=\overline
\newcommand{\be}{\begin{equation}}
\newcommand{\ee}{\end{equation}}
\def\ft#1#2{{\textstyle{{\scriptstyle #1}\over {\scriptstyle #2}}}}
\def\fft#1#2{{#1 \over #2}}
\def\vp{\varphi}
\def\sst#1{{\scriptscriptstyle #1}}
\def\oneone{\rlap 1\mkern4mu{\rm l}}
\def\td{\tilde}
\def\wtd{\widetilde}
\def\ie{\rm i.e.\ }
\def\dalemb#1#2{{\vbox{\hrule height .#2pt
        \hbox{\vrule width.#2pt height#1pt \kern#1pt
                \vrule width.#2pt}
        \hrule height.#2pt}}}
\def\square{\mathord{\dalemb{6.8}{7}\hbox{\hskip1pt}}}
\def\wtd{\widetilde}
\def\R{\rlap{\rm I}\mkern3mu{\rm R}}
\def\im{{\rm i}}
\def\tilg{\tilde{g}}
\def\tilF{\tilde{F}}
\def\tilA{\tilde{A}}
\def\varf{\varphi}
\def\tilf{\tilde{\phi}}
\def\tilh{\tilde{h}}
\def\rme{{\rm e}}
\def\te{\tilde{e}}
\def\tG{\tilde{G}}
\def\ep{\varepsilon}
\newcommand{\1}{{(1)}}
\newcommand{\2}{{(2)}}
\newcommand{\3}{{(3)}}
\newcommand{\4}{{(4)}}
\newcommand{\5}{{(5)}}
\newcommand{\xx}{{\underline{x}}}
\newcommand{\hD}{\hat{D}}
\newcommand{\hC}{\hat{C}}
\newcommand{\hg}{\hat{g}}
\newcommand{\hj}{\hat{j}}
\newcommand{\hB}{\hat{B}}
\newcommand{\hcB}{\hat{{\cal B}}}
\newcommand{\hA}{\hat{A}}
\newcommand{\hf}{\hat{\phi}}
\newcommand{\hvf}{\hat{\varphi}}
\newcommand{\hchi}{\hat{\chi}}
\newcommand{\umu}{{\underline{\mu}}}
\newcommand{\unu}{{\underline{\nu}}}
\newcommand{\urho}{{\underline{\rho}}}
\newcommand{\usigma}{{\underline{\sigma}}}
\newcommand{\ur}{{\underline{r}}}
\newcommand{\ua}{{\underline{a}}}
\newcommand{\ub}{{\underline{b}}}
\newcommand{\uc}{{\underline{c}}}
\newcommand{\ud}{{\underline{d}}}
\newcommand{\ualpha}{{\underline{\alpha}}}
\newcommand{\ubeta}{{\underline{\beta}}}
\newcommand{\ugamma}{{\underline{\gamma}}}
\newcommand{\udelta}{{\underline{\delta}}}
\usepackage{graphicx}% Include figure files
\usepackage{dcolumn}% Align table columns on decimal point
\usepackage{bm}% bold math
\begin{document}
\preprint{MIT-CTP-3575}
\title{Rotating Brane Worlds and the Global Rotation of the Universe}
\author{Alan H. Guth$^1$}%
 \email{guth@ctp.mit.edu}
\author{Ali Nayeri$^{1,2}$}%
 \email{nayeri@MIT.EDU}
\affiliation{$^1$~Center for Theoretical Physics, Laboratory for
Nuclear Science and Department
of Physics,\\
Massachusetts Institute of Technology, Cambridge,
Massachusetts 02139\\
$^2$~Institute for Fundamental Theory, Department of Physics,
University of Florida, Gainesville, Florida 32611}
\date{December 9, 2004}
\begin{abstract}
We introduce a class of brane-world models in which a single
brane is embedded in an anti-de Sitter spacetime containing a
rotating (Kerr) black hole.  In this Letter we consider the case
of slow rotation, calculating the metric and dynamics of the
brane world to first order in the angular momentum of the black
hole.  To this order we find that the cosmic fluid on the brane
rotates rigidly relative to a Robertson-Walker frame of
reference, which in turn rotates rigidly relative to the original
Kerr-anti-de Sitter coordinate frame.  Corrections to the
Friedmann equations and the shape of the brane occur only at
higher order.  We construct models for which the geometry on the
brane is either closed or open, but the open models are described
only for small distances from the rotation axis, and may very
likely develop pathologies at larger distances.
\end{abstract}
\pacs{11.25.-w, 11.25.Wx, 11.27.+d, 98.80.Jk}

\maketitle In recent years, there has been a great deal of interest
in studying the cosmology of universes with extra dimensions
\cite{ADD:1998,RS1:1999,RS2:1999}. The common feature of all these
models is the distinction of the observable universe (the brane
world) from the rest of the universe (the bulk). While all the gauge
fields with spin less than two are confined to the brane, the
gravitons can propagate into the bulk.  The cosmology of brane-world
models is in general different from standard
Friedmann-Robertson-Walker (FRW) cosmology, but if the bulk is
anti-de Sitter space (AdS), containing perhaps a Schwarzschild black
hole, then the cosmology is very similar to FRW at late times
\cite{BDEL:2000}.  The presence of a static black hole in the bulk
shows up in the modified Friedmann equation as {\it ``dark
radiation''} \cite{KRAUS:1999,IDA:2000,AAA:2001}. Though rotating
black holes in higher dimensions \cite{MP:1986} and with asymptotic
AdS behavior in four dimensions \cite{Carter:1968} were studied many
years ago, it was after the advent of the AdS/CFT correspondence
\cite{Maldacena:1998} that the properties of Kerr-AdS black holes in
higher dimensions were studied in great detail
\cite{Klemm:1998,Hawking:1999}.  The purpose of this Letter is to
investigate the behavior of a brane world in the presence of a bulk
Kerr-AdS$_5$ black hole in the slowly rotating regime. In this
five-dimensional spacetime the rotating black hole can be
characterized by two independent projections of the angular
momentum, $J_\varphi$, and, $J_\psi$, but in this Letter we consider
only the case in which $J_\psi = 0$, with $J_\varphi \equiv j$.  The
stationary axisymmetric metric for a five-dimensional
single-parameter Kerr black hole can be written as
%%%%%%%%%%%%%%%%%%%%%%%%%%
\ba ds^2_5  &=&  - \frac{\Delta_r}{\rho^2} \left[d T -
\frac{k\,j}{\Xi_k}\, S_k^2(\theta)\, d\varphi\right]^2 +
\frac{\rho^2}{\Delta_r}d r^2 + \frac{\rho^2}{\Delta_\theta} d
\theta^2 \nn \\
&& + \frac{\Delta_\theta\,S^2_k(\theta)}{\rho^2}\left[k^2\,j\, d T
- \frac{(k^2\, j^2 + r^2)}{\Xi_k}\, d\varphi\right]^2 \nn \\
&& + r^2 C_k^2(\theta)\, d \psi^2\,,   \label{full_kerr_ads}\ea
%%%%%%%%%%%%%%%%%%%%%%%%%%%%%%%%%
where
%%%%%%%%%%%%%%%%%%%%%%%%%%%%%%%%%%%
\ba \Delta_r & = & (k^2\,j_{\varphi}^2 + r^2)\left(k +
\frac{r^2}{\ell^2} \right)
- 2m \,,\nn \\
\rho^2 & = & r^2 + k^2\,j^2 \, C_k^2(\theta) \,, \nn \\
\Xi_k & = & 1 - k \,\frac{j^2}{\ell^2}\,, \nn \\
\Delta_\theta & = & 1 - k\,\frac{j^2}{\ell^2}
C_k^2(\theta)\,,
 \ea
%%%%%%%%%%%%%%%%%%%%%%%%%%%%%%%%%%%
and
%%%%%%%%%%%%%%%%%%%%%%%%%%%%%%%%%%%%%%%
\ba S_k(\theta) & = &\frac{\sin{(\sqrt{k}\,\theta)}}{\sqrt{k}} \nn \\
C_k(\theta)& = & \frac{d S_k(\theta)}{d \theta} =
\cos{({\sqrt{k}\,\theta})} \,.\ea
%%%%%%%%%%%%%%%%%%%%%%%%%%%%%%%%%%%%%%
The parameter $m>0$ is related to the mass of the black hole, $j$
to the angular momentum for rotation in the $\varphi$ direction,
$\ell^2 = -6/\Lambda_5$ is the AdS$_5$ length scale, where
$\Lambda_5$ is the cosmological constant, and $k$ is the
curvature parameter.  A change in $|k|$ with no change in the
sign of $k$ can be compensated by a rescaling of $r$, $\theta$,
$\psi$, $T$, $m$, and $j$, but a change in the sign of $k$
results in a change in the topology of the hypersurfaces of
constant $r$ and $T$.

Note that as $k \rightarrow 0$, $j$ completely disappears from
the metric, so there is no rotation in this limit.  As far as we
know there is no solution that behaves smoothly as $k \rightarrow
0$ and which preserves a nontrivial effect of $j$.  A $k = 0$
solution with nontrivial $j$ as a distinct case has been
described by Klemm \cite{Klemm:1998}.  The metric
(\ref{full_kerr_ads}) for $k=1$ was given by Hawking et
al.~\cite{Hawking:1999}.

We will confine our discussion to values of $r$ that are large
enough to ensure that $\Delta_r(r)>0$, and for $k>0$ we also
require $j^2 < \ell^2/k$ to ensure that the metric has the proper
signature.  With these restrictions, we can show that the
spacetime does not contain any closed timelike or null curves.
To see this, note that any closed curve can be parameterized by
$X^A(\lambda)$, where $X^A \equiv (T, r, \theta, \varphi, \psi)$
and $\lambda \in [0, 1]$ is a parameter, with $dX^A/d \lambda \ne
0$ for all $\lambda$. The curve must have a maximal value of
$T(\lambda)$, at which $d T/d\lambda=0$.  At this point $(dX^A/d
\lambda)^2$ contains only terms proportional to $dr^2$, $d
\theta^2$, $d \psi^2$, and $d \varphi^2$, where the first 3 are
manifestly positive.  With some algebra the coefficient of $d
\varphi^2$ can be written as
%%%%%%%%%%%%%%%%%%%%%%%%%%%%%%%%%%
\be
g_{\varphi\varphi} = \frac{S_k^2(\theta)}{\Xi_k^2 \rho^2} \left[
(r^2 + k^2 j^2) \rho^2 \Xi_k + 2 m k^2 j^2 S_k^2(\theta) \right]
\,, \ee
%%%%%%%%%%%%%%%%%%%%%%%%%%%%%%%%%%
which is also positive.  Thus $(dX^A/d \lambda)^2 > 0$, and
therefore the curve cannot be timelike or null.

To study the dynamics of a brane world in this background, we
simplify the problem by working perturbatively through first
order in the rotation parameter $j$.  In this approximation the
metric (\ref{full_kerr_ads}) may be written as
%%%%%%%%%%%%%%%%%%%%%%%%%%%%%%%%%%
\ba d s^2_5 & \thickapprox & - F_k(r)\,d T^2 + \frac{d
r^2}{F_k(r)} \nn \\ & &+ 2\, kj \, S_k^2(\theta) \, G_k(r) \, d T
\, d \varphi + r^2 d\Omega_3^2 \,, \label{kerr-ads-metric} \ea
%%%%%%%%%%%%%%%%%%%%%%%%%%%%%%%%
where
%%%%%%%%%%%%%%%%%%%%%%%%%%%%%%%%
\be G_k(r) = - \frac{2m}{r^2} + \frac{r^2}{\ell^2}\,, \quad
F_k(r) =  G_k(r)+k\,, \label{Fdef} \ee
%%%%%%%%%%%%%%%%%%%%%%%%%%%%%%%%%%
and
%%%%%%%%%%%%%%%%%%%%%%%%%%%%%%%%%
 \be
 d \Omega_3^2 =
 d \theta^2 + S_k^2(\theta)\, d\varphi^2 +
 C_k^2(\theta)\, d \psi^2\,.
 \ee
%%%%%%%%%%%%%%%%%%%%%%%%%%%%%%%
For $k>0$ this linearization is valid throughout the manifold,
for sufficiently small $j$, but for $k<0$ the unboundedness of
$S_k^2(\theta)$ and $C_k^2(\theta)$ implies that for any $j$ the
linearization will break down for sufficiently large $\theta$.

Suppose that a brane is located at $r = r_b(T)$, for some
function $r_b(T)$, and that we construct a
$\mathbb{Z}_2$-symmetric spacetime consisting of the region $r <
r_b(T)$ of the original Kerr-AdS metric (\ref{full_kerr_ads}),
with the spacetime for $r > r_b(T)$ replaced by a mirror copy of
the spacetime for $r < r_b(T)$.  The position of the brane can be
conveniently described by $N (X^A) = 0$, where $N = r - r_b(T)$.
The spacetime of the brane can be described in 4D coordinates
$x^\mu \equiv (t, \theta, \varphi, \psi)$, with an induced metric
%%%%%%%%%%%%%%%%%%%%%%%%%%%%%%%%%
\ba d s^2_4& = & g_{AB} \frac{\partial X^A}{\partial x^\mu} \,
\frac{\partial X^B}{\partial x^\nu} \, d x^\mu \, d x^\nu \equiv
g_{AB} \, Q^A{}_{\mu} \, Q^B{}_\nu \, d x^\mu \, d x^\nu \nn \\
& \equiv &
\gamma_{\mu\nu} \, d x^\mu \, d x^\nu \nn \\
& = & - d t^2 + 2 \,j k \, S^2_k(\theta) \, G_k(a) \left(\frac{d
T}{d t}\right) \, d t \, d \varphi + a^2 \, d \Omega_3^2\,,\nn \\
\label{4metric_orginal} \ea
%%%%%%%%%%%%%%%%%%%%%%%%%%%%%%%%%%%%%%%%%%%%%
where
%%%%%%%%%%%%%%%%%%%%%%%%%%%%%%%%%%%%%%%%%%%%%%%
\be d t = \left[F_k(a) - \frac{1}{F_k(a)} \left(\frac{d  r_b}{d
T}\right)^2\right]^{1/2} \, d T\,, \ee
%%%%%%%%%%%%%%%%%%%%%%%%%%%%%%%%%%%%%%%%%%%%%%%%%
and the Robertson-Walker (RW) scale factor is given by $a(t) =
r_b\bigl(T(t)\bigr)$.  Using an overdot to denote a derivative
with respect to $t$, the above relation can be rewritten as
%%%%%%%%%%%%%%%%%%%%%%%
\be \frac{d  T}{d t} = \frac{\sqrt{F_k(a) + \dot
a^2}}{F_k(a)}\,.\ee
%%%%%%%%%%%%%%%%%%%%%%%%%%%
To first order in $j$, the metric (\ref{4metric_orginal}) can be
diagonalized to give a Robertson-Walker metric by introducing a
new angular coordinate $\phi$ defined by
%%%%%%%%%%%%%%%%%
\be \phi(\varphi,t) = \varphi -\int^t \Omega(t')\,d
t'\,, \label{RWphi} \ee
%%%%%%%%%%%%%%%%%%
where
%%%%%%%%%%%%%%%%%%
\be \Omega(t) = - k j\left[\frac{G_k(a) \, \sqrt{F_k(a)+\dot
a^2}}{a^2(t) \, F_k(a)}\right]\,.\ee
%%%%%%%%%%%%%%%%%%
$\Omega(t)$ is thus the angular velocity of the RW frame with
respect to the Kerr-AdS$_5$ frame. In the RW coordinates the
metric is simply
%%%%%%%%%%%%%%%%%%%%%%%%%%
\be d s^2_4 = - d t^2 + a^2(t) \, d \Omega_3'^2
\,,\label{4metric_diagonal} \ee
%%%%%%%%%%%%%%%%%%%%%%%%%
where
%%%%%%%%%%%%%%%%%%%%%%%%%%
\be d \Omega_3'^2 =  d \theta^2 + S^2_k(\theta)\, d \phi^2 +
C^2_k(\theta)\, d \psi^2\,.\ee
%%%%%%%%%%%%%%%%%%%%%%%%%%%

We can now study the dynamics of the brane through the junction
conditions on the brane.  To this aim, one needs to calculate the
extrinsic curvature $K_{\mu\nu}$, which is defined in terms of
the normalized outward normal vector
%%%%%%%%%%%%%%%%%%%%%%%%%%%%%%%%
\be n_A = \frac{N_{,A}}{\sqrt{g^{BC}N_{,B} N_{,C}}} = - \dot a\,
\delta_A\,^T + \frac{\sqrt{F_k(a) + \dot a^2}}{F_k(a)}\,
\delta_A\,^r \,. \ee
%%%%%%%%%%%%%%%%%%%%%
(We will use commas to denote coordinate derivatives, and
semicolons to denote covariant derivatives.)  The extrinsic
curvature is defined by
%%%%%%%%%%%%%%%%%%%%%
\be K_{\mu\nu} \equiv n_{B;A} \, Q^A{}_\mu \, Q^B{}_\nu \,,\ee
%%%%%%%%%%%%%%%%%%%%%
and to first order in $j$ the nonzero components are found to be
%%%%%%%%%%%%%%%%%%%%%%%%%%%%%%%%%%%%%%%%%%%%%%%%%%%%%%%%%%
\ba K^t{}_t &=& \frac{a}{\sqrt{F_k(a)+ H^2 a^2}} \left(\dot H
+ H^2 + \frac{1}{\ell^2}+\frac{2m}{a^4} \right) \,, \nn \\
K^t{}_\varphi &=& -4 \, k j \, m S_k^2(\theta)/a^3 \,,
\nn \\
K^\varphi{}_t &=& - \frac{k j}{a^3 F_k(a)}\left[a^2 G_k(a) \dot
H - k \left( \frac{a^2}{\ell^2} + \frac{2m}{a^2} \right)\right]
\,, \nn \\
K^\theta{}_\theta &=& K^\varphi{}_\varphi = K^\psi{}_\psi = \sqrt{
F_k(a) + H^2 a^2}/a \,, \ea
%%%%%%%%%%%%%%%%%%%%%%%%%%%%%%%%%%%%%%%%%%%%%%%%%%%%
where $H \equiv \dot a/a$ is the Hubble parameter on the brane.
The extrinsic curvature can be related to the energy momentum
tensor $S^{\mu\nu}(x)$ on the brane \footnote{The full 5D energy
momentum tensor will contain a term $\tilde S^{AB}\delta(\eta)$,
where $\eta$ is a normalized coordinate orthogonal to the brane,
and $S_{\mu\nu}(x) \equiv S_{AB}\, Q^A{}_\mu \, Q^B{}_\nu$.}
through the junction condition \cite{Israel:1966}.  Assuming the
$\mathbb{Z}_2$ symmetry on the brane, the junction condition with
the sign appropriate to keeping the region $r < r_b(T)$ may be
written as
%%%%%%%%%%%%%%%%%%%%%%%%%%%%%%%%%%%%%%%%%%%%%%
\be S^\mu\,_\nu = \frac{2}{\kappa^2_5} [K^\mu\,_\nu -
\delta^\mu\,_\nu K] \,,\ee
%%%%%%%%%%%%%%%%%%%%%%%%%%%%%%%%%%%%%%%%%%%%%%%
where $K = \gamma^{\mu\nu} K_{\mu\nu}$ is the trace of the
extrinsic curvature.  Straightforward calculation leads to
%%%%%%%%%%%%%%%%%%%%%%%%%%%%%%%%%%%%%%%%%%%%%%%
\ba
K &=& \frac{a}{\sqrt{F_k(a)+H^2 a^2}} \nn \\
&&\qquad \times \left(\dot H+4 H^2 + \frac{4}{\ell^2}-\frac{4
m}{a^4} + \frac{3 k}{a^2} \right) \,, \nn \\
%%%%%%%%%%%%%%%%%%%%%%%%%%%%%%%%%%%%%%%%%%%%%%%
S^t{}_t &=& - \frac{6 \sqrt{F_k(a)+H^2 a^2}}{\kappa_5^2
a} \label{s_tau_tau} \nn \,,\\
%%%%%%%%%%%%%%%%%%%%%%%%%%%%%%%%%%%%%%%%%%%%%%%
S^t{}_\varphi &=& - \frac{8 \, k j \, m}{\kappa_5^2 a^3}\, S_k^2
(\theta) \label{s_tau_phi} \nn \,,\\
%%%%%%%%%%%%%%%%%%%%%%%%%%%%%%%%%%%%%%%%%%%%%%%
S^\varphi{}_t &=& - \frac{2 k j}{\kappa_5^2 a^3 F_k(a)}
\left[a^2 \, G_k(a) \, \dot H - k \left( \frac{a^2}{\ell^2} +
\frac{2 m}{a^2} \right) \right] \,, \nn \\
\label{s_phi_tau} \\
%%%%%%%%%%%%%%%%%%%%%%%%%%%%%%%%%%%%%%%%%%%%%%%
\noalign{\noindent and finally}
%%%%%%%%%%%%%%%%%%%%%%%%%%%%%%%%%%%%%%%%%%%%%%%
S^\theta{}_\theta &=& S^\varphi{}_\varphi = S^\psi{}_\psi =
-\frac{2 a}{\kappa_5^2 \sqrt{F_k(a)+H^2 a^2}} \nn \\
&& \quad \times \left(\dot H+3 H^2 + \frac{3}{\ell^2} - \frac{2
m}{a^4} + \frac{2 k}{a^2} \right) \label{s_angular} \,.\ea
%%%%%%%%%%%%%%%%%%%%%%%%%%%%%%%%%%%%%%%%%%%%%%%

Now suppose that $S_{\mu\nu}$ consists of two distinct parts,
namely,
%%%%%%%%%%%%%%%%%%%%%%%%%%%%%%%%%%%%%%%%%%%%%%%%%%%%%%
\be S^{\mu\nu} = -\lambda \gamma^{\mu\nu} +
\tau^{\mu\nu}\,,\label{braneenergytensor}\ee
%%%%%%%%%%%%%%%%%%%%%%%%%%%%%%%%%%%%%%%%%%
where $\lambda$ is the tension of the brane and
$\tau^{\mu\nu}$ is the energy-momentum tensor of the matter fluid
on the brane.  If the matter can be described as a perfect fluid,
then
%%%%%%%%%%%%%%%%%%%%%%%%%%%%%%%%%%%%%%%%%
\be \tau^\mu{}_\nu = (\rho_m + p_m) u^{\mu}\,u_{\nu} +
p_m\,\delta^\mu_\nu \,, \label{brane_EMT}\ee
%%%%%%%%%%%%%%%%%%%%%%%%%%%%%%%%%%%%%%%%%%%
in which case $S^\mu{}_\nu$ has one eigenvector with eigenvalue
$-(\rho_m+\lambda)$, and three degenerate eigenvectors with
eigenvalue $p_m - \lambda$.  To linear order in $j$,
$S^\mu{}_\nu$ as shown above has exactly this pattern of
eigenvalues, so we can treat it as a perfect fluid.  The
eigenvalues (to this order) are given by the diagonal elements,
so we see immediately that
%%%%%%%%%%%%%%%%%%%%%%%%%%%%%%%%%%%%%%%%%%%
\be \rho_m  + \lambda = - S^t{}_t = \frac{6
\sqrt{F_k(a)+H^2 a^2}}{\kappa_5^2 a} \,, \label{rhoeq} \ee
%%%%%%%%%%%%%%%%%%%%%%%%%%%%%%%%%%%%%%%%%%%
which with Eq.~(\ref{Fdef}) can be rewritten as a generalized
Friedmann equation,
%%%%%%%%%%%%%%%%%%%%%%%%%%%%%%%%%%%%%%%%%%%
\be H^2 = \frac{1}{3} (8 \pi G_N \rho_m + \Lambda_4) +
\frac{\kappa_5^4}{36} \rho_m^2 -
\frac{k}{a^2} + \frac{2m}{a^4} \, , \ee
%%%%%%%%%%%%%%%%%%%%%%%%%%%%%%%%%%%%%%%%%%%
where $G_N = \lambda \kappa_5^4 /(48 \pi)$ and $\Lambda_4 =
\frac{1}{2} (\Lambda_5 + \frac{1}{6} \kappa_5^4 \lambda^2)$.  To
this order $j$ does not enter the Friedmann equation, so the
above equation is the standard one for brane-world cosmology with
a Schwarzschild black hole \cite{KRAUS:1999,IDA:2000,AAA:2001}.
Similarly $p_m - \lambda$ is the 3-fold degenerate eigenvalue, so
$p_m - \lambda = S^\theta{}_\theta = S^\varphi{}_\varphi =
S^\psi{}_\psi$ as given in Eq.~(\ref{s_angular}).  This
expression looks complicated, but with Eq.~(\ref{rhoeq}) it
implies that
%%%%%%%%%%%%%%%%%%%%%%%%%%%%%%%%%%%%%%%%%%%
\be \rho_m + p_m = - \frac{2a}{\kappa_5^2 \sqrt{F_k(a) + H^2 a^2}}
\left( \dot H - \frac{k}{a^2} + \frac{4 m}{a^4} \right) \,,
\label{rhop} \ee
%%%%%%%%%%%%%%%%%%%%%%%%%%%%%%%%%%%%%%%%%%%
which by using Eq.~(\ref{rhoeq}) again can be shown to be
equivalent to the standard energy conservation equation,
%%%%%%%%%%%%%%%%%%%%
\be \dot \rho_m + 3 H (\rho_m + p_m) = 0  \,. \ee
%%%%%%%%%%%%%%%%%%%%%%%%%%%%%%%%%%%%

The fluid velocity $u^\mu$ is determined by the fact that it is
the eigenvector of $S^\mu{}_\nu$ with eigenvalue $S^t{}_t$.
Although the eigenvalue is not affected by $j$ to first order,
the eigenvector is affected, being given by $u^\theta = u^\psi =
0$, with
%%%%%%%%%%%%%%%%%%%%%%%%%%%%%%%%%%%%%%%%%%%
\ba u^\varphi &=& \frac{S^\varphi{}_t}{S^t{}_t -
S^\varphi{}_\varphi} \, u^t
= \frac{k j \, \sqrt{F_k(a) + H^2 a^2}}{a^2} \nn \\
&& \quad \times \left[ \frac{4 m}{a^4 \dot H - a^2 k + 4 m} -
\frac{G_k(a)}{F_k(a)} \right] \, u^t \,. \label{KerrVel}
\ea
%%%%%%%%%%%%%%%%%%%%%%%%%%%%%%%%%%%%%%%%%%%
Since $u^\varphi/u^t = d \varphi/ d t$ is a function only of $t$,
the above equation describes rigid rotation, in which all points
move with the same angular velocity.

Eq.~(\ref{KerrVel}) shows that the cosmic fluid is rotating
rigidly with respect to the Kerr-AdS$_5$ frame of reference, but
observers on the brane would have no way of measuring this
quantity.  We can, however, transform this result to the RW frame
defined by Eq.~(\ref{RWphi}), which implies that
%%%%%%%%%%%%%%%%%%%%%%%%%%%%%%%%%%%%%%%%%%%
\be \frac{d \phi}{d t} = \frac{d \varphi}{d t} - \Omega(t)
= \frac{4 \, k j \, m \sqrt{F_k(a) + H^2 a^2}}{a^6 \left(\dot H
- \frac{k}{a^2} + \frac{4 m}{a^4} \right)} \,. \label{DphiDt} \ee
%%%%%%%%%%%%%%%%%%%%%%%%%%%%%%%%%%%%%%%%%%%
Since the galaxies would on average be comoving with the cosmic
fluid, the angular velocity given above would be directly
measurable as an apparent rotation of the distant galaxies
relative to the locally inertial (RW) frame of reference.

To observers on the brane, the rotation of the distant galaxies
relative to the RW frame would be interpreted as a clear
violation not only of Mach's principle, but also the Einstein
equations, since in this frame the Einstein tensor is diagonal
but the energy-momentum tensor is not.  In brane-world cosmology
the Einstein equations are in general modified \cite{SMS:2000} by
nonlinear terms and by adding to $8 \pi G_N \tau_{\mu\nu}$ the
term $-E_{\mu\nu}$, where
%%%%%%%%%%%%%%%%%%%%%%%%%%%%%%%%%%%%%%%%%%%
\be
E_{\mu\nu} \equiv {}^{(5)}C^A{}_{B C D}\, n_A \, n^C \, Q^B{}_\mu
\, Q^D{}_\nu \ ,
\ee
%%%%%%%%%%%%%%%%%%%%%%%%%%%%%%%%%%%%%%%%%%%
and ${}^{(5)}C^A{}_{B C D}$ is the 5-dimensional Weyl tensor.
To lowest order in $j$, the nonzero components of $E_{\mu\nu}$
in the RW frame (with sign conventions of Ref.~\cite{SMS:2000})
are given by
%%%%%%%%%%%%%%%%%%%%%%%%%%%%%%%%%%%%%%%%%%%
\ba
E_{\tau\tau} &=& \frac{3 E_{\theta\theta}}{a^2} = \frac{3
E_{\phi\phi}}{a^2 S_k^2(\theta)} = \frac{3 E_{\psi\psi}}{a^2
C_k^2(\theta)} = - \frac{6 m}{a^4} \nn \\
E_{\phi\tau} &=& E_{\tau\phi} = \frac{8 k j m \sqrt{F_k(a)+H^2
a^2}}{a^4} \, S_k^2(\theta) \ .
\ea
%%%%%%%%%%%%%%%%%%%%%%%%%%%%%%%%%%%%%%%%%%%
Thus, the diagonal entries of $E_{\mu\nu}$ are responsible for the
{\it ``dark radiation''}, and the off-diagonal entries are
responsible for what might be called {\it ``dark angular
momentum''}. In the RW frame this {\it ``dark angular momentum''}
cancels the physical angular momentum, so the metric can remain
isotropic.

Eq.~(\ref{DphiDt}) expresses $d \phi / d t$ in terms of the
kinematical quantities $H$ and $\dot H$, but we can also express
the result in terms of $\rho$ and $p$.  By combining
Eqs.~(\ref{rhop}) and (\ref{DphiDt}), one has
%%%%%%%%%%%%%%%%%%%%%%%%%%%%%%%%%%%%%%%%%%%
\be \frac{d \phi}{d t}=-\frac{8\, k j \, m}{\kappa_5^2 a^5
(\rho_m + p_m)} \,. \label{DphiDt2} \ee
%%%%%%%%%%%%%%%%%%%%%%%%%%%%%%%%%%%%%%%%%%%
In this form we can tell how the rotational velocity of the
universe would vary with time.  The quantity $\rho_m + p_m$ would
in most situations be dominated by either radiation or
nonrelativistic matter, since a cosmological constant would make
no contribution.  In the former case $d \phi/d t$ falls off as
$1/a$, and in the latter it falls off as $1/a^2$.  Thus the
rotation would be strongly suppressed by inflation, during which
$a$ would grow by many orders of magnitude.  Nonetheless, in
models with minimal inflation and large initial angular momenta,
it is not impossible for the universe to have an observable
rotation rate in the present era.

Finally, to verify the properties of the rotation in a manifestly
coordinate-invariant way, we calculate the vorticity and the
shear of the fluid velocity.  The vorticity and the shear are
defined in terms of $\theta_{\alpha\beta} \equiv u_{\mu;\nu}
q^\mu{}_\alpha q^\nu{}_\beta$, where $q_{\mu\nu} \equiv
\gamma_{\mu\nu} + u_\mu u_\nu$ is the projection tensor
orthogonal to the four-velocity $u^\mu$.  The vorticity is the
antisymmetric part, $\omega_{\alpha \beta} \equiv
\theta_{[\alpha,\beta]}$ and $\omega^2 \equiv \frac{1}{2}
\omega_{\alpha\beta} \omega^{\alpha\beta}$.  The shear is the
traceless symmetric part, $\sigma_{\alpha\beta} \equiv
\theta_{(\alpha,\gamma)} - \frac{1}{3} q_{\alpha\beta}
\theta^\gamma_\gamma$, with $\sigma^2 \equiv \frac{1}{2}
\sigma_{\alpha\beta}\,\sigma^{\alpha\beta}$.  To evaluate these
quantities, one first normalizes the velocity vector described in
Eq.~(\ref{KerrVel}), which to lowest order requires $u^t=1$.
Lowering the index using $\gamma_{\mu\nu}$ then gives $u_\theta =
u_\psi = 0$, $u_t = -1$, and

% For some reason the following equation did not TeX properly
%   with the current ArXiv processor:
%     TeX, Version 3.14159 (Web2C 7.3.1)
%     LaTeX2e <1999/12/01> patch level 1
%     revtex4 2001/08/03 v4.0
% It works normally on other systems, such as
%     TeX, Version 3.14159 (MiKTeX 2.1)
%     LaTeX2e <2000/06/01>
%     revtex4 2001/08/03 v4.0
% The problem is that the TeX processor fails to include
%   \parfillskip after the ``and'' before the equation, and then
%   inserts extra space before the equation.  In this case the
%   extra space causes the paper to stretch onto a 5th page.  I
%   do not know what is special about this text, since the ArXiv
%   processor certainly spaces text correctly in most
%   circumstances.
% My (Alan Guth's) fix is to insert a paragraph break, cancelling
%   the extra space with the following skips:
\prevdepth=0pt \vskip -\baselineskip \vskip -\parskip
%%%%%%%%%%%%%%%%%%%%%%%%%%%%%%%%%%%%%%%%%%%
\be u_\varphi = - \frac{8 \, k j \, m}{\kappa_5^2 \, a^3 \, (\rho_m +
p_m)} S_k^2(\theta) \,, \ee
%%%%%%%%%%%%%%%%%%%%%%%%%%%%%%%%%%%%%%%%%%%
where we again used Eq.~(\ref{rhop}).  One then finds that the
shear vanishes identically, as expected for rigid rotation, and
the vorticity has the nonzero components
%%%%%%%%%%%%%%%%%%%%%%%%%%%%%%%%%%%%%%%%%%%
\be \omega_{\theta\varphi} = -\omega_{\varphi\theta} = \frac{8
\,k\, j \, m}{\kappa_5^2\, a^3 (\rho_m + p_m)}
S_k(\theta)\,C_k(\theta) + {\cal O}(j^2) \label{vorticity} \,,\ee
%%%%%%%%%%%%%%%%%%%%%%%%%%%%%%%%%%%%%%%%%%%
giving
%%%%%%%%%%%%%%%%%%%%%%%%%%%%%%%%%%%%%%%%%%%
\be \omega^2 = \frac{64 j^2 k^2 m^2 }{\kappa_5^4 a^{10}
(\rho_m + p_m)^2} \, C_k^2(\theta) + {\cal O}(j^3)
\, . \ee
%%%%%%%%%%%%%%%%%%%%%%%%%%%%%%%%%%%%%%%%%%%
Note that $\omega^2$ can be written as $\omega^2 = (d \phi / d
t)^2 C_k^2(\theta)$, where $d \phi/ d t$ was given in
Eq.~(\ref{DphiDt2}).  Thus, in the vicinity of the axis of
rotation, the vorticity is just the square of the angular
velocity relative to the locally inertial frame, as one would
expect from Newtonian physics.

The main result of this paper is that a brane-world spacetime
with a Kerr-AdS$_5$ black hole in the bulk provides a very simple
model of a universe exhibiting global rotation:  if the angular
momentum is small, the matter in the brane world rotates rigidly
relative to the inertial frame of reference.  It would be
interesting to calculate the implications of such a model for the
cosmic background radiation.  The rigid rotation is perfectly
consistent for a closed brane universe, but for an open brane
universe such rigid rotation would lead at large distances (large
$\theta$) to spacelike velocities.  Our linearized approximation
would break down at such large distances, but we do not know if a
consistent model can be constructed.  These and other issues are
under investigation.

We wish to thank Yuk-Yan Lam and David Poland for many useful
discussions we had in the early stage of preparing this paper.
A.H.G. was supported in part by funds provided by the U.S.
Department of Energy (D.O.E.) under cooperative research
agreement \#DF-FC02-94ER40818.


\begin{thebibliography}{23}

\bibitem{ADD:1998} N.~Arkani-Hamed, S.~Dimopoulos, and G. Dvali,
Phys.~Lett.~{\bf B429}, 263 (1998).

\bibitem{RS1:1999} L.~Randall and R.~Sundrum,
Phys.~Rev.~Lett.~{\bf 83}, 3370 (1999).

\bibitem{RS2:1999} L.~Randall and R.~Sundrum,
Phys.~Rev.~Lett.~{\bf 83}, 4690 (1999).

\bibitem{BDEL:2000} P. Bin\'{e}truy, C. Deffayet, E. Ellwanger,
and D.~Langlois, Phys.~Lett.~{\bf B477}, 285 (2000).

\bibitem{KRAUS:1999} P.~Kraus, J.~High~Energy~Phys.~{\bf 9912},
011 (1999).

\bibitem{IDA:2000} D.~Ida, J.~High~Energy~Phys.~{\bf 0009}, 014
(2000).

\bibitem{AAA:2001} A.~Chamblin, A. Karch and A.~Nayeri,
Phys.~Lett.~{\bf B509} 163 (2001).

\bibitem{MP:1986} R.~Myers and M.~Perry, Ann.~Phys.~{\bf 172},
304 (1986).

\bibitem{Carter:1968} B.~Carter, Commun.~Math.~Phys.~{\bf 10},
280 (1968).

\bibitem{Maldacena:1998} J.~Maldacena, Adv.~Theor.~Math.~Phys.~{\bf 2},
231 (1998).

\bibitem{Klemm:1998} D.~Klemm, J.~High~Energy~Phys.~{\bf 9811},
019 (1998).

\bibitem{Hawking:1999} S. W. Hawking, C. J. Hunter, and M. M.
Taylor-Robinson, Phys. Rev. D {\bf 59}, 064005 (1999).

\bibitem{Israel:1966} W.~Israel, Nuovo Cim.\ B {\bf 44}, 1 (1966)
[Erratum: Ibid.\ B {\bf 48}, 463 (1967)].

\bibitem{SMS:2000} T.~Shiromizu, K.~Maeda, and M.~Sasaki,
Phys.~Rev.~D~{\bf 62}, 024012 (2000).

\end{thebibliography}
\end{document}